# Heavy Anionic Complex Creates a Unique Water Structure at a Soft Charged Interface

*William Rock,[†] Baofu Qiao,[†] Tiecheng Zhou,[‡] Aurora E. Clark,[‡] and Ahmet Uysal[†,*]*

[†] Chemical Sciences and Engineering Division, Argonne National Laboratory, Argonne, IL 60439, United States

[‡] Department of Chemistry and the Material Science and Engineering Program, Washington State University, Pullman, WA 99164, United States

AUTHOR INFORMATION

**Corresponding Author**

*E-mail: ahmet@anl.gov. Web: www.ahmet-uysal.com. Phone: +1-630-252-9133

ABSTRACT

Ion hydration and interfacial water play crucial roles in numerous phenomena ranging from biological to industrial systems. Although biologically relevant (and mostly smaller) ions have been studied extensively in this context, very little experimental data exist about molecular scale behavior of heavy ions and their complexes at interfaces, especially under technologically significant conditions. It has recently been shown that $PtCl_6^{2-}$ complexes adsorb at positively charged interfaces in a two-step process that cannot fit into well-known empirical trends, such as Hofmeister series. Here, a combined vibrational sum frequency generation and molecular dynamics study reveals that a unique interfacial water structure is connected to this peculiar adsorption behavior. A novel sub-ensemble analysis of MD simulation results show that after adsorption, $PtCl_6^{2-}$ complexes partially retain their first and second hydration spheres, and it is possible to identify three different types of water molecules around them based on their orientational structures and hydrogen bonding strengths. These results have important implications for relating interfacial water structure and hydration enthalpy to the general understanding of specific ion effects. This in turn influences interpretation of heavy metal ion distribution across and reactivity within, liquid interfaces.

**INTRODUCTION**

The adsorption of ions at aqueous charged interfaces drives many macroscopic processes including chemical separations,[1] desalination,[2] protein solvation,[3-4] atmospheric chemistry,[5-7] and geochemistry.[8-9] Specific ion effects make modeling these processes difficult; a thorough understanding of the molecular-scale interactions that govern them is necessary.[10] Here, vibrational sum frequency generation (VSFG) spectra reveal a novel interfacial water structure due to the adsorption of $PtCl_6^{2-}$ complexes at a soft charged interface. MD simulations show that this structure is due to $PtCl_6^{2-}$ partially retaining its hydration sphere after closely adsorbing to the interface. This results is in stark contrast to the trends observed in Hofmeister anions, where easily shedding the hydration sphere is correlated with a high surface affinity. A molecular-scale description of this unusual water structure is obtained through detailed sub-ensemble analysis of MD simulation, and offers a new approach to the analysis of VSFG spectra.

Electrostatics usually control the interaction of charged particles with surfaces; however, when a hydrated ion is close to a charged interface, short-range weak interactions become important.[11] The interfacial chemistry is further complicated by the fact that the aqueous medium is not a uniform continuum, but contains considerable structural organization.[12] It is known that deviations from ideal solvent behavior are related to specific ion effects.[13-14] Although recent experimental,[11, 15-17] computational,[18] and theoretical[19-20] advances have improved our understanding considerably, the emerging view is that specific ion effects may not have a single unified explanation, especially when ion-ion interactions and interfacial ion-binding sites are operative.[17]

The interfacial interactions of biologically relevant, and usually lighter, ions in aqueous environments have been extensively studied.[15-16, 21-22] Yet comparable work on heavier elements, including the platinum group metals (PGMs), lanthanides, and actinides, are scarce.[23] This is despite the fact that many modern lighting, displays, and carbon-free energy technologies rely on the efficient refining and reprocessing of heavy metals based upon liquid:liquid extraction (LLE) technologies;[1, 24] most of these processes involve the interaction of aqueous interfaces with the anionic complexes of heavy elements in highly concentrated solutions.[24-26]

In addition to its practical implications, a molecular scale understanding of interfacial water in the presence of large ions is the best way to test the limits of the theories that explain specific-ion effects. It is known that the thermochemical radii of ions plays an important role in their interfacial interactions,[23] and a recent study demonstrated qualitative differences in the adsorption behavior of heavy and light anions at a charged interface.[27] However, the surface x-ray scattering techniques used in that study did not provide any information about the interfacial water structure. The current work leverages the well-established ability of VSFG to study interfacial water-ion interactions in numerous systems[4, 16, 21-22, 28-30] (including Langmuir monolayers on electrolyte solutions - Figure 1), which allows for facile control of the interactions between organic functional groups and ions.[4, 23, 29] Using this approach we provide clear evidence, a unique signal in VSFG spectra at 3600 cm$^{-1}$ appearing as a result of adsorbed ions, that heavy anions can adsorb tightly to a charged interface and remain strongly hydrated, indicating that adsorption behavior does not always follow hydration strength. The main experiments are designed to keep all parameters constant except the bulk concentration of $PtCl_6^{2-}$, to make the interpretation of the data relatively simple. The sub-ensemble analysis of MD simulations, and other supportive experiments also independently strengthen this interpretation.

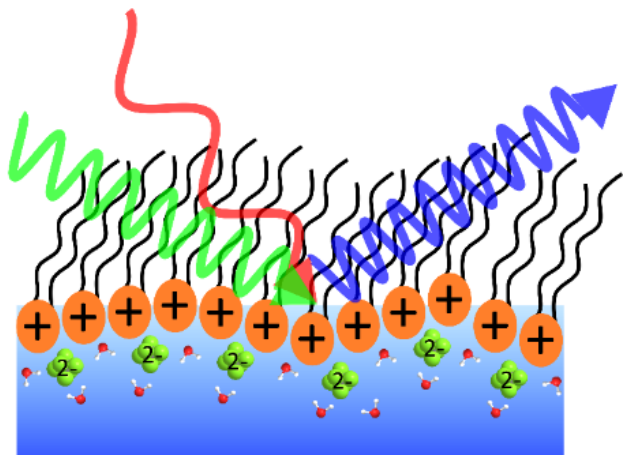

**Figure 1.** A schematic of VSFG measurements at the air/water/DPTAP interface. Tunable IR and fixed 532 nm laser pulses overlap spatially and temporally at the interface to generate VSFG signal

**EXPERIMENTAL AND COMPUTATIONAL METHODS**

**VSFG Setup.** Experiments were done with an EKSPLA VSFG spectrometer. The VSFG setup begins with a picosecond laser (PL2231-50) containing a master oscillator with a Nd:YVO$_4$ laser rod and a regenerative amplifier using a diode-pumped Nd:YAG rod. The amplified 1064 nm output is ~28 mJ with a pulse width of ~29 ps and a 50 Hz repetition rate. A Harmonics Unit (SFGH500-H/2H) splits the laser output into three paths; KDP crystals frequency-double two paths. The 1064 nm path and one of the 532 nm paths pump an OPA/DFG (PG501-DFG1P) using BBO/AgGaS$_2$ crystals that produces narrowband infrared (IR) pulses tunable from ~1000-4000 cm$^{-1}$. The 532 nm polarization is adjusted with a λ/2 waveplate, and the VSFG signal is selected using a Glan polarizer. The IR polarization is selected by adjusting a pair of motorized mirrors that control the beam path through a periscope. After the sample, the visible and IR excitations are blocked with an iris, and are further attenuated with a 532 nm notch filter and a 510 nm shortpass filter. The SFG signal is then directed to a monochromator (Sol, MS2001) and collected with a PMT (Hamamatsu, R7899).

The VSFG spectrometer employs a reflection geometry; the visible and IR excitation angles, with respect to the sample normal, are θ$_{vis}$ = 60º and θ$_{IR}$ = 55º, respectively. The visible and IR energies at the sample are ~800 µJ and ~65 µJ, respectively. A motorized piezoelectric rotation stage (ThorLabs, ELL8K) rotates the sample after each frequency step to avoid sample damage. Each spectrum was collected with a 4 cm$^{-1}$ increment over the range 2800-3800 cm$^{-1}$, and averaged 300 laser shots per point. The spectra were normalized against the SFG spectrum of z-cut quartz. Both the SSP and SPS polarization conditions were collected for each sample. PPP polarization was also collected for the highest concentration (not shown) sample and consistent with the interpretation presented below. The polarization conditions are defined using the industry standard, S(S)- VSFG signal, S(P)- VIS excitation, and P(S)- IR excitation. The electric field of P-polarized light is parallel to the plane of incidence, and the electric field of S-polarized (from the German senkrecht) light is perpendicular to the plane of incidence.

**VSFG Data Global Fit.** The AIWHB band cannot be fit by a single Lorentzian (Supporting Information). Therefore, it is globally fit to a two-Lorentzian model, which suggests that at least two different water environments contribute strongly to the AIWHB band:

$$\left| A_{NR} e^{i\varphi_{NR}} + \frac{A_{1n}}{\omega - \omega_1 - i\Gamma_1} + \frac{A_{2n}}{\omega - \omega_2 - i\Gamma_2} \right|^2 \quad (1)$$

The nonresonant amplitude and phase ($A_{NR}, \varphi_{NR}$), the peak location ($\omega_1, \omega_2$), and the peak width ($\Gamma_1, \Gamma_2$) are linked in the global fit, and the peak amplitudes ($A_{1n}, A_{2n}$) are allowed to float in each spectrum. A high-quality two-peak global fit requires one positive and one negative peak; however, which of the peaks is positive or negative depends on the initial guess (the absolute phase cannot be determined from this experiment). The imaginary part of the fit is given by the following equation:

$$Im(\chi_r^2) = \frac{A_{1n}\Gamma_1}{(\omega-\omega_1)^2+\Gamma_1^2} + \frac{A_{2n}\Gamma_2}{(\omega-\omega_2)^2+\Gamma_2^2} \qquad (2)$$

The peaks were found to be centered at $\omega_1$=3534 cm$^{-1}$ and $\omega_2$=3606 cm$^{-1}$. The SPS intensity is much stronger than the SSP intensity in the 3534 cm$^{-1}$ peak, and SPS and SSP intensities are approximately equal in the 3606 cm$^{-1}$ peak. The fit parameters and the comparison of the various fits based on different models can be found in the supporting information.

**FTIR Measurements.** The spectra were collected with a Nicolet Nexus 870 FTIR spectrometer with an attenuated total reflectance accessory.

**Sample Preparation.** Anhydrous lithium chloride (LiCl, 99%), chloroplatinic acid solution (H$_2$PtCl$_6$, 8 wt. % in H$_2$O), and HPLC grade chloroform (CHCl$_3$, ≥ 99.9%) were purchased from Sigma-Aldrich. 1 N hydrochloric acid (HCl) was purchased from Fisher Scientific. 1,2-dioleoyl-3-trimethylammonium-propane (DPTAP) chloride salt was purchased in powder form from Avanti Polar Lipids and stored at -20° C. All chemicals were used as received. Aliquots of 0.25 mM DPTAP in CHCl$_3$ were prepared and stored at -20° C. DPTAP aliquots are discarded in ≤ 5 days and are not subject to any freeze-thaw cycles. All subphase solutions are 20 mL, contain 500 mM LiCl, and are adjusted to pH 2 using 1 N HCl. The 5 mM PtCl$_6^{2-}$ solution is at pH ~2 without the addition of HCl. Ultrapure water with a resistivity of 18.2 MΩcm (Barnstead, Nanopure TOC-UV) was used to prepare each subphase.

Langmuir monolayer samples are prepared in a 60x20 mm flat-form PTFE dish. A Nima pressure sensor (from a model 601A Langmuir trough) using a chromatography paper Wilhelmy plate measures the surface pressure of the Langmuir experiments are performed at room-temperature (~20° C) and at a surface pressure between 10-13 mN/m. The Langmuir monolayer is prepared using drop-wise addition of 0.25 mM DPTAP in CHCl$_3$ from a 1 µL Hamilton syringe. A similar system has been investigated in a Langmuir trough under fixed pressure with synchrotron x-ray scattering techniques.[27] Those measurements suggest that the DPTAP monolayer structure is stable under these conditions for the duration of the experiments. The SFG measurements of the CH region for each sample in this study (not shown) support this expectation.

**Water Solvation Sub-ensemble Analysis.** The water solvation sub-ensemble analysis has been performed on the trajectories of previously reported classical molecular dynamics (MD) simulations,[27] which used the GROMACS package (version 4.5.5)[31] and the CHARMM 36 force field.[32-34] GROMACS is a generic molecular dynamics (MD) simulation package for molecular structure at the nanometer resolution. GROMACS has been extensively employed in studying proteins, surfactants, lipid bilayers, and similar systems.[35-37] GROMACS has also been employed in studying liquid/air and solid/liquid surfaces.[38-39]

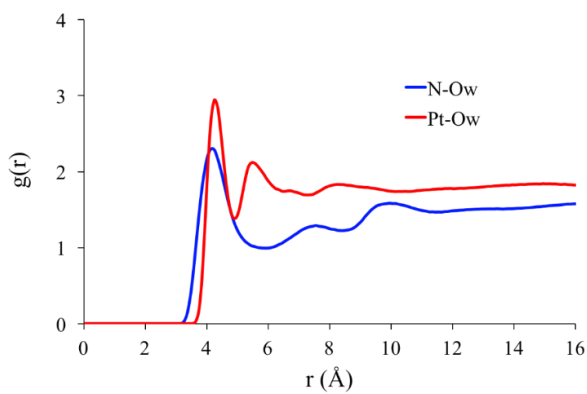

**Figure 2**. (red) RDF between the metal center (Pt atom) and water (Ow atom), and (blue) RDF between the surfactant head group (N atom) and water (Ow atom)

The simulations were done at full $PtCl_6^{2-}$ surface coverage with 0.5 M LiCl background to represent the highest concentration experiments. The molecular area per DPTAP molecule in the simulations was set to 48 $Å^2$, as determined by grazing incidence X-ray diffraction measurements.[27] The radial distribution functions in Figure 2 are calculated from the MD simulations in order to determine the solvation environment of $PtCl_6^{2-}$ and DPTAP surfactant head group. The first peak of the Pt-Ow RDF at r ~ 4.25 Å corresponds to the first solvation shell of $PtCl_6^{2-}$, with a first minimum at r ~ 5.00 Å. Thus the first solvation shell of $PtCl_6^{2-}$ can be defined as $r_{Ow..Pt} < 5.00$ Å. The second peak of the Pt-Ow RDF at r ~ 5.45 Å corresponds to the second solvation shell of $PtCl_6^{2-}$, with a second minimum at r ~ 6.50 Å. Therefore, the second solvation shell of $PtCl_6^{2-}$ can be defined as $5.00 < r_{Ow..Pt} < 6.50$ Å. Similarly, in Figure 2 the first peak in the N-Ow RDF occurs at r ~ 4.15 Å with its minimum at r ~ 6.00 Å, corresponding to the solvation shell of DPTAP surfactant head group. Therefore, the solvation shell of DPTAP head group can be defined as $r_{Ow..N} < 6.00$ Å.

Considering the solvation environment of water around the metallate and surfactant head group, as shown in the radial distribution functions in Figure 2, we decompose the water molecules in our simulation box into four groups:

**AIWHB-Up:** water molecules in the 1$^{st}$ solvation shell of $PtCl_6^{2-}$, i.e. $r_{Ow..Pt} < 5.00$ Å;

**AIWHB-Down:** water molecules in the 2$^{nd}$ solvation shell of $PtCl_6^{2-}$, and also within the solvation shell of the DPTAP head group, i.e. $5.00 < r_{Ow..Pt} < 6.50$ Å and $r_{Ow..N} < 6.00$ Å;

**WWHB-Down:** water molecules outside the 2$^{nd}$ solvation shell of $PtCl_6^{2-}$ and within the solvation shell of the DPTAP head group, i.e. $r_{Ow..Pt} > 5.00$ Å and $r_{Ow..N} < 6.00$ Å;

**WWHB-Up:** water molecules outside the 1$^{st}$ solvation shell of $PtCl_6^{2-}$ and also outside the solvation shell of the DPTAP head group, i.e. $r_{Ow..Pt} > 5.00$ Å and $r_{Ow..N} > 6.00$ Å

**RESULTS AND DISCUSSION**

**Observation of Anion Induced Interfacial Water Structure.** A typical VSFG spectrum at the neat air/water interface has two significant features: a sharp peak at 3700 cm$^{-1}$ corresponding to the free OH stretch, and a broad band between ~3050-3500 cm$^{-1}$ corresponding to the OH modes of water:water hydrogen bonds (WWHB).[16] Spreading a charged Langmuir monolayer on pure water leads to the disappearance of the free OH and to significant enhancement of the WWHB signal.[28, 40] When ions are introduced to the subphase, the WWHB signal decreases or completely disappears depending on effect of the ion.[21, 41-44] These effects usually correlate with the Hofmeister series and hydration strength. Larger, more polarizable ions easily shed their hydration shells and strongly associate with charged interfaces. For instance, the affinity of halide anions for the positively charged 1,2-dioleoyl-3-trimethylammonium-propane (DPTAP) interface follows the Hofmeister series;[45] larger I$^-$ anions adsorb close enough to completely eliminate the WWHB VSFG signal. This is usually interpreted as the loss of orientational order of interfacial water.[15, 45] While many different salts and acids have been similarly investigated, the observed effects of these ions were limited solely to the changes in the intensity and shape of the WWHB peak.[16] In this study, VSFG spectroscopy, MD simulations and sub-ensemble analysis reveal a unique interfacial water structure after the adsorption of $PtCl_6^{2-}$ to a DPTAP monolayer. One aspect of this unusual structure is the emergence of a weakly hydrogen-bonded interfacial water band at ~3600 cm$^{-1}$. The water VSFG signal at 3600 cm$^{-1}$, caused by the presence of interfacial ions, *has not* been previously observed. In addition, the emergence of the 3600 cm$^{-1}$ signal coincides with the disappearance of the WWHB signal (~3050-3500 cm$^{-1}$). In contrast to the straightforward interpretation – a loss of orientation order – sub-ensemble analysis of MD simulation trajectories reveals that, in this system, the vanishing VSFG signal is a result of oppositely oriented regions of ordered water molecules.

Figure 3a presents the VSFG spectra of the air/water interface and the air/DPTAP/5000 μM PtCl$_6^{2-}$ solution interface from 2800-3800 cm$^{-1}$ taken using SSP polarization. VSFG spectra are collected using the SSP and SPS polarization conditions (see Experimental Details). All PtCl$_6^{2-}$ subphase solutions contain 500 mM LiCl and are adjusted to pH 2 using HCl to mimic industrial separations conditions. This also helps to fix the PtCl$_6^{2-}$ speciation and control the PtCl$_6^{2-}$ adsorption at the interface as a function of the bulk concentration. Henceforth, the solutions will be delineated using only the [PtCl$_6^{2-}$]. The air/water VSFG spectrum is included as a benchmark.[30, 46-48]

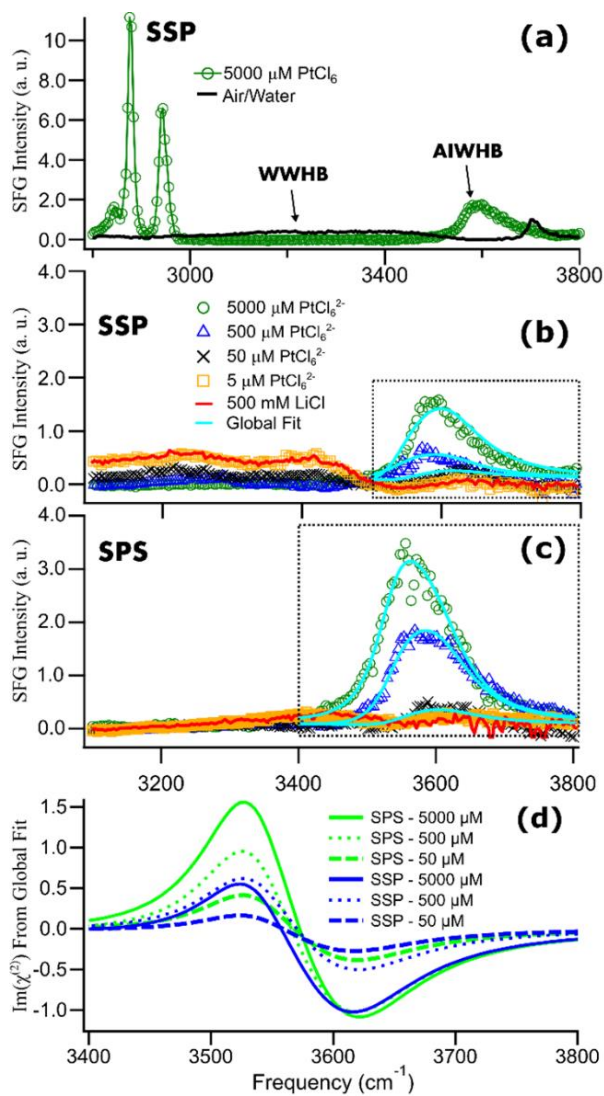

**Figure 3.** VSFG spectra of a DPTAP monolayer on $PtCl_6^{2-}$ solutions. (a) Full acquired spectral range (2800-3800 cm$^{-1}$) including the CH and OH stretching regions compared to the air/water interface (b, c) SSP and SPS VSFG spectra, respectively, of the OH stretching region (3100-3800 cm$^{-1}$). Cyan lines are a global fit to the weakly hydrogen-bonded region – marked with dashed boxes – of the 3 highest [$PtCl_6^{2-}$] spectra. (d) The imaginary part of the global fit.

Figures 3b and 3c zoom in on the OH-stretching region (3100-3800 cm$^{-1}$) of the air/DPTAP/$PtCl_6^{2-}$ solution spectra taken using the SSP and SPS polarization conditions, respectively. The spectra span three orders of magnitude of $PtCl_6^{2-}$ concentration (5 to 5000 μM), and include a background spectrum without $PtCl_6^{2-}$ (500 mM LiCl at pH 2). The air/DPTAP/water VSFG spectrum has been reported previously.[40, 45]

The VSFG spectrum of DPTAP on 500 mM LiCl is comparable to previous results at lower concentrations,[45] showing some WWHB signal, suggesting that $Cl^-$ partially shields the interfacial water molecules from the DPTAP surface charge. Also, the monolayer blocks the free-OH groups, eliminating the narrow 3700 cm$^{-1}$ peak. The addition of 5000 μM $PtCl_6^{2-}$ eliminates the WWHB signal, and gives rise to a new band around 3600 cm$^{-1}$. The intensity and structure of the new band depends on the $[PtCl_6^{2-}]$. Hydrogen-bonding red-shifts the water OH vibration (from ~3700 cm$^{-1}$ in free-OH to ~3050-3500 cm$^{-1}$ in WWHB); since the magnitude of the red-shift depends on the strength of the intermolecular interaction, this new peak is due to a weakly hydrogen-bonded water structure. Although some previous water studies have reported weakly hydrogen-bonded peaks around 3600 cm$^{-1}$ due to water molecules trapped in the hydrophobic tail regions of Langmuir monolayers,[29, 49-52] this is the first evidence of an *anion-induced weakly hydrogen-bonded* (AIWHB) interfacial water structure.

Although previous studies attribute water stretches around 3600 cm$^{-1}$ to water molecules interacting with the hydrophobic tails of Langmuir monolayers, that is likely not the case here. A previous surface x-ray study on the same system shows that the area per DPTAP molecule (at a constant surface pressure) decreases with increasing $[PtCl_6^{2-}]$.[27] Therefore, the addition of $PtCl_6^{2-}$ should not increase the amount of water molecules between the lipid tails. In addition, as it is explained in the next section in detail, our MD simulations show that only a very little amount of water molecules are present above the DPTAP head groups, and all of these water molecules are in the hydration shell of the $PtCl_6^{2-}$.

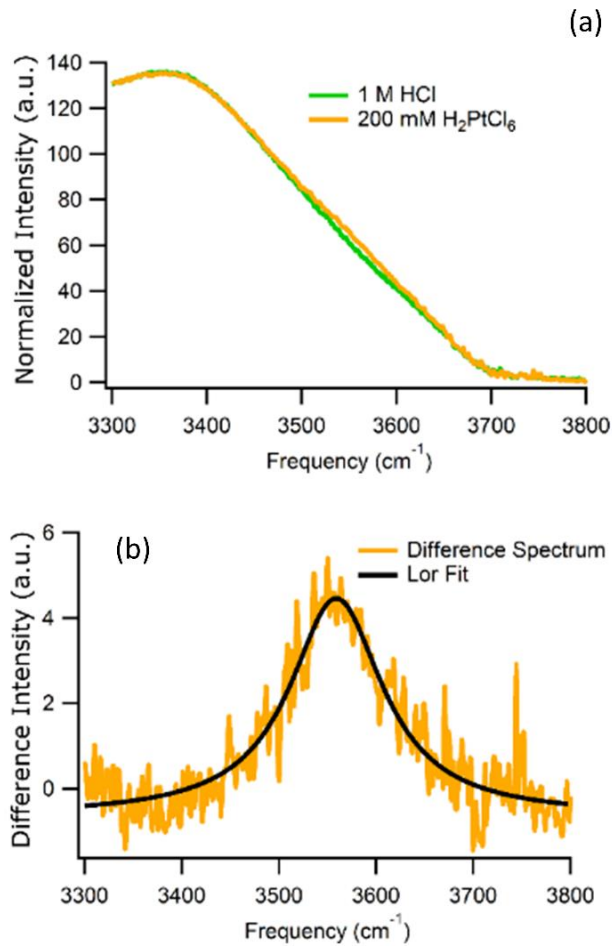

**Figure 4.** (a) FTIR spectra of 1 M HCl and 200 mM $H_2PtCl_6$ stock solutions. (b) Difference of the two spectra in (a) along with a Lorentzian fit. The fit is centered at 3559 cm$^{-1}$.

ATR-FTIR measurements provide further evidence that the AIWHB band arises from water solvating $PtCl_6^{2-}$ anions. Figure 4a shows FTIR spectra of 1 M HCl and 200 mM $H_2PtCl_6$ stock solutions. The spectrum of the highly concentrated stock solution is collected because the effect of $PtCl_6^{2-}$ anions on the bulk water spectrum is very small. The spectra are dominated by bulk water signal; however, the difference of the two spectra shows a peak at 3559 cm$^{-1}$. This peak is likely due to water solvating $PtCl_6^{2-}$, and is fit with a single Lorentzian (Figure 4b). This supports the conclusion that the VSFG band at ~3600 cm$^{-1}$ is due to water molecules solvating $PtCl_6^{2-}$. The positively charged DPTAP interface distorts the electronic structure of $PtCl_6^{2-}$ and induces multiple water sub-ensembles, which causes the VSFG peaks to shift and induces the appearance of a band rather than a single peak. This is also in agreement with previous infrared pump-probe measurements that found the O-H---Cl⁻ stretch centered at 3450 cm$^{-1}$ and the O-H---I⁻ stretch centered at 3500 cm$^{-1}$.[53] It is reasonable to expect a smaller red-shift for larger, more polarizable $PtCl_6^{2-}$ anions.

The AIWHB band can be best fit by two oppositely signed Lorentzians (Figure 3d), which suggests that the spherically symmetric hydration shell in the bulk is changed at the interface and at least two different water environments contribute strongly to the AIWHB band. The peaks were found to be centered at $\omega_1$=3534 cm$^{-1}$ and $\omega_2$=3606 cm$^{-1}$. The SPS intensity is much stronger than the SSP intensity in the 3534 cm$^{-1}$ peak, and SPS and SSP intensities are approximately equal in the 3606 cm$^{-1}$ peak. The sign of the peaks cannot be determined directly without heterodyne detected VSFG (HD-VSFG) measurements, and is instead obtained by sub-ensemble analysis of MD simulation.

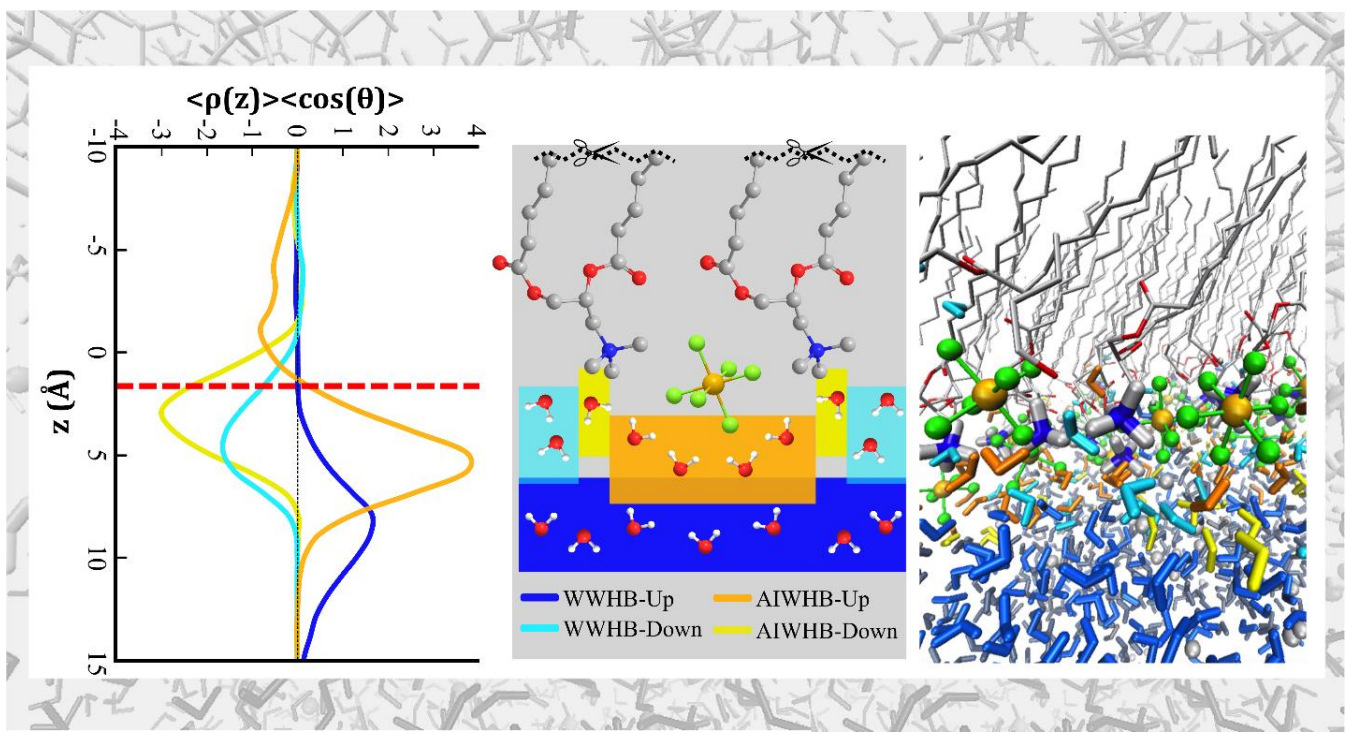

**Figure 5.** Weighted dipole orientation plot (left), a simplified representation of the water structures MD predicts (middle), an actual MD snapshot of the differently oriented waters (right). The plot (left) is the product of the average water number density ($<\rho(z)>$) and the average water orientation ($<\cos(\theta)>$) – where $\theta$ is the angle between the water dipole vector and the surface normal – vs the z-dimension coordinate (z(Å)). z(0) is the average position of the DPTAP N atoms, and the red dashed line is the average Pt position. MD plots 4 water sub-ensembles: AIWHB-up (orange) and AIWHB-down (yellow) waters hydrogen bond to $PtCl_6^{2-}$. WWHB-up (blue) and WWHB-down (cyan) waters hydrogen bond to other water molecules (Experimental details). The color scheme is matched across all 3 panels. The cartoon (middle) and snapshot (right) approximately align with the plot z – the shaded boxes (middle) are set to the FWHM of the MD plots. The water sub-ensembles are named after the predominant hydrogen-bond-donor character (WWHB or AIWHB) and the average H-atom orientation (Up or Down).

**Water Solvation Sub-ensemble Analysis.** We next analyze the dipole orientation and water organization with MD simulations. The product $<\rho(z)><\cos(\theta)>$, the average water number density weighted by its orientation, has been shown to be a useful estimate of VSFG intensity.[54-55] The orientational angle ($\theta$) of individual $H_2O$ molecules is defined as the angle that the dipole moment vector of $H_2O$ makes with the surface normal. Figure 5 presents $<\rho(z)><\cos(\theta)>$ of four different interfacial water environments versus the z-dimension; the gas phase is on the top (z < -10 Å), and the water bulk is on the bottom (z > 15 Å). The total $<\rho(z)><\cos(\theta)>$ may obscure structural details if there are multiple water environments (with different vibrational frequencies) at the same distance (z) from the interface. As such, we decomposed the total $<\rho(z)><\cos(\theta)>$ into sub-ensembles of different intermolecular interactions. This approach is a good alternative to calculating the VSFG spectra explicitly using MD simulations, which is resource intensive and requires parameter optimization for the non-ideal solutions under study.[56]

Four primary water environments were ascertained based on their nearest possible interactions (either the $N^+$ of DPTAP or $PtCl_6^{2-}$) and explained in Figure 5. A negative $<\rho(z)><\cos(\theta)>$ indicates that water is oriented with its H-atoms pointing away from the surface (H-down), and a positive value indicates that the water-hydrogens point towards the surface (H-up); this is the same sign convention used in $Im(\chi^{(2)})$ in Figure 3d.

The MD simulations predict that the first hydration shell of $PtCl_6^{2-}$ has ~10 water molecules in the bulk, and this number drops to ~5 at the interface.[27] These AIWHB-up waters mainly stay underneath the $PtCl_6^{2-}$ ions (Figure 5) and give rise to the positive peak in Figure 3d. Also a very little amount of water is observed to remain above $PtCl_6^{2-}$ which create the small and broad negative peak at z<0. The second hydration shell also exhibits an asymmetric structure at the interface. AIWHB-down waters are defined as being in the second hydration shell of $PtCl_6^{2-}$ *and* in the first hydration shell of DPTAP $N^+$. Therefore, they are closer to the interface, where the first hydration shell of $PtCl_6^{2-}$ is less dense (as can be seen by comparing the AIWHB-up (orange) and AIWHB-down (yellow) curves in Figure 5). This allows AIWHB-down water molecules to hydrogen bond to $PtCl_6^{2-}$, but with a weaker intermolecular interaction than AIWHB-up waters in the first hydration shell. (The first and second hydration shells are defined by water distances from $PtCl_6^{2-}$, as explained in experimental details). These AIWHB-down waters are responsible for the less red shifted negative peak in Figure 3d. The rest of the second hydration shell is found in the WWHB-up sub-ensemble, underneath the $PtCl_6^{2-}$. These waters feel the negative charge from $PtCl_6^{2-}$ above and adopt an H-up orientation, hydrogen bonding to the water molecules in the first hydration shell, and should have a peak in the WWHB region. However, WWHB-down waters – which are outside the 2$^{nd}$ solvation shell of $PtCl_6^{2-}$ *but within* the solvation shell of $N^+$, and adopt an H-down orientation due to the positive charge from DPTAP $N^+$ – create an equal but opposite signal in WWHB region. Therefore, these oppositely oriented sub-ensembles cancel and no WWHB signal is observed in the VSFG spectrum. This is in stark contrast to other systems, such as I$^-$, in which the WWHB peak disappears because the strongly adsorbed ions destroy the orientational ordering of water molecules.[45]

**Detailed Investigation of AIWHB Peak.** The interpretation above depends on the assumption that the two peaks in the AIWHB band in Figure 3 actually originate from waters in two different environments, and not from a splitting of a single peak due to vibrational coupling, as is observed in the VSFG spectra of the air/water interface.[57] We repeated the 5mM $PtCl_6^{2-}$ measurement in a 50:50 volume % of $H_2O$:$D_2O$ to address this issue (Figure 6). The intensity of the AIWHB band decreased approximately by a factor of two as expected, but the spectral shape did not change, supporting the premise that couplings do not complicate the lineshape of the AIWHB band. In Figure 6, the AIWHB band is also observed to appear in the absence of 500 mM LiCl, providing further evidence that it is a result of $PtCl_6^{2-}$ adsorption.

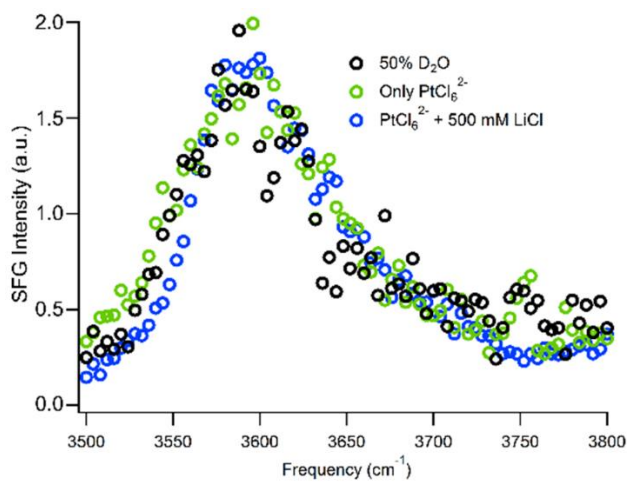

**Figure 6.** SSP VSFG spectra of DPTAP with a surface pressure of ~10 mN/m on three different subphases: 5 mM $H_2PtCl_6$/500 mM LiCl in $H_2O$ (blue), 5 mM $H_2PtCl_6$/500 mM LiCl in 50:50 $H_2O$:$D_2O$ (black), and 5 mM $H_2PtCl_6$ in $H_2O$. The intensity is scaled to match the 5 mM $H_2PtCl_6$/500 mM LiCl in $H_2O$ spectrum to show that the peak shape remains the same.

Further details about the nature of the AIWHB band can be obtained from polarization dependent experiments. Wang and coworkers developed a framework to deduce the type of vibrational mode and molecular orientation angle from polarization dependent VSFG spectra.[58-59] Interfacial water OH stretches have $C_{2v}$ or $C_{\infty v}$ symmetry depending on whether their hydrogen atoms are in a symmetric or asymmetric environment, respectively.[58-59] Regardless of the molecular orientation angle, if water adopts $C_{\infty v}$ symmetry, the SSP spectra will be more intense than the SPS spectra, and if water adopts $C_{2v}$ symmetry, the asymmetric stretch will be much stronger in the SPS spectra, and the symmetric stretch will be much stronger in the SSP spectra. Therefore, the SPS/SSP intensity ratios of the AIWHB band suggest that it predominantly arises from the asymmetric stretch of water with $C_{2v}$ symmetry. The asymmetric water stretch has previously been observed in Langmuir monolayer systems, and was attributed to water molecules interacting with the lipid tails [51]; although this is not the first observation of the asymmetric water stretch, the fact that it is observed suggests that the adsorbed ions create a unique water environment, because the WWHB stretch at the air/water interface is dominated by the symmetric stretch.[58]

Here, we presented a detailed investigation of AIWHB peak with all the tools available to us at the moment. Nevertheless, further investigations of AIWHB peak with HD-VSFG will be important to elucidate all aspects of its orientational structure, since the phase information is lost in the presented results. It is also important to expand these studies to other heavy anionic complexes, such as $PdCl_4^{2-}$, to see how AIWHB peak depends on the detailed structure of the complex.

**CONCLUSIONS**

The results presented here provide evidence of an unexpected orientational structure of interfacial water due to adsorbed $PtCl_6^{2-}$ ions with important implications to both the role of the interfacial water molecules in amphiphile-ion interactions, and their detection through vibrational spectroscopy. Hofmeister trends are often explained using hydration energies – ions that easily shed their hydration shell adsorb at interfaces much more easily than strongly hydrated ions.[14, 42-45] However, $PtCl_6^{2-}$ anions both adsorb tightly to the DPTAP interface and are strongly hydrated.[27] The present work demonstrates that $PtCl_6^{2-}$ retains part of its hydration shell after adsorption and can keep the second hydration shell oriented (WWHB-Up), a behavior that is showing that larger multivalent anions behave significantly differently than their lighter Hofmeister anions.

This system also leads to an interesting interfacial water structure with a complex profile of positively and negatively charged patches (Figure 5). Averaging these structures within the excitation volume, as is done in VSFG measurements, may lead to the loss of information. The most common reason for the absence of VSFG signal is a lack of interfacial orientational order. However, these results highlight the fact that the absence of VSFG signal can also indicate the presence of oppositely ordered sub-ensembles. Therefore, integrating MD simulation and sub-ensemble analysis with VSFG experiments is an important approach for interpretation.

This newly observed interfacial water structure may play a crucial role in chemical processes (i.e. LLE) involving heavy metal anions, which usually rely on a few $k_BT$ of free energy difference between different phases for interfacial ion transfer.[60] Interfacial water structures and hydration effects may create energetic barriers comparable to such small free energy differences. For instance, it has recently been shown that the competitive adsorption of $PtCl_6^{2-}$ ions on amine functionalized silicon surfaces[61] and at DPTAP monolayers[27] do not follow the predictions of mean field theories, and the differences are caused by hydration effects and ion-ion correlations. In a broader perspective, these experiments show that the generalizations that are based on lighter and simpler ions cannot be easily extended to predict the behavior of heavier and more complex ions at interfaces, and specific experiments, combined with appropriate computational tools, are necessary to address them.

**ASSOCIATED CONTENT**

The supporting information is available free of charge on the ACS Publications website.

Details of the VSFG Data global fit and determining the interfacial hydrogen bonding environment; wider range VSFG spectrum of $D_2O$ dilution studies.

**AUTHOR INFORMATION**

**Corresponding Author**

*E-mail: ahmet@anl.gov. Web: www.ahmet-uysal.com. Phone: +1-630-252-9133

**Notes**

The authors declare no competing financial interest.

**ACKNOWLEDGMENTS**

We thank Lynda Soderholm for her valuable comments, and Richard Wilson for FTIR measurements. This work is supported by the U.S. Department of Energy, Office of Basic Energy Science, Division of Chemical Sciences, Geosciences, and Biosciences, under contract DE-AC02-06CH11357. T.Z. and A.E.C. are supported by the same program under contract DE-SC0001815. The MD simulations were done at the computing resources provided on Blues, a high-performance computing cluster operated by the Laboratory Computing Resource Center at Argonne National Laboratory.

**REFERENCES**

1. Sholl, D. S.; Lively, R. P. Seven chemical separations to change the world. *Nature* **2016,** *532* (7600), 435-437.
2. Rica, R. A.; Ziano, R.; Salerno, D.; Mantegazza, F.; Brogioli, D. Thermodynamic relation between voltage-concentration dependence and salt adsorption in electrochemical cells. *Phys. Rev. Lett.* **2012,** *109* (15), 156103.
3. Lo Nostro, P.; Ninham, B. W. Hofmeister phenomena: an update on ion specificity in biology. *Chem. Rev.* **2012,** *112* (4), 2286-322.
4. Sung, W.; Avazbaeva, Z.; Kim, D. Salt promotes protonation of amine groups at air/water interface. *J. Phys. Chem. Lett.* **2017,** *8* (15), 3601-3606.

5. Casillas-Ituarte, N. N.; Callahan, K. M.; Tang, C. Y.; Chen, X.; Roeselova, M.; Tobias, D. J.; Allen, H. C. Surface organization of aqueous MgCl2 and application to atmospheric marine aerosol chemistry. *Proc. Natl. Acad. Sci. USA* **2010,** *107* (15), 6616-21.
6. Hua, W.; Verreault, D.; Allen, H. C. Surface prevalence of perchlorate anions at the air/aqueous interface. *J. Phys. Chem. Lett.* **2013,** *4* (24), 4231-4236.
7. Hua, W.; Jubb, A. M.; Allen, H. C. Electric field reversal of Na2SO4, (NH4)2SO4, and Na2CO3 relative to CaCl2 and NaCl at the air/aqueous interface revealed by heterodyne detected phase-sensitive sum frequency. *J. Phys. Chem. Lett.* **2011,** *2* (20), 2515-2520.
8. Fenter, P.; Lee, S. S. Hydration layer structure at solid-water interfaces. *Mrs. Bull.* **2014,** *39* (12), 1056-1061.
9. DeWalt-Kerian, E. L.; Kim, S.; Azam, M. S.; Zeng, H.; Liu, Q.; Gibbs, J. M. pH-dependent inversion of Hofmeister trends in the water structure of the electrical double layer. *J. Phys. Chem. Lett.* **2017,** *8* (13), 2855-2861.
10. Mazzini, V.; Craig, V. S. J. What is the fundamental ion-specific series for anions and cations? Ion specificity in standard partial molar volumes of electrolytes and electrostriction in water and non-aqueous solvents. *Chemical Science* **2017,** *8* (10), 7052-7065.
11. Wang, W. J.; Park, R. Y.; Travesset, A.; Vaknin, D. Ion-specific induced charges at aqueous soft interfaces. *Phys. Rev. Lett.* **2011,** *106* (5), 056102.
12. Marcus, Y. Effect of ions on the structure of water: structure making and breaking. *Chem. Rev.* **2009,** *109* (3), 1346-70.
13. McCaffrey, D. L.; Nguyen, S. C.; Cox, S. J.; Weller, H.; Alivisatos, A. P.; Geissler, P. L.; Saykally, R. J. Mechanism of ion adsorption to aqueous interfaces: Graphene/water vs. air/water. *Proc. Natl. Acad. Sci. USA* **2017,** *114* (51), 13369-13373.
14. Okur, H. I.; Hladilkova, J.; Rembert, K. B.; Cho, Y.; Heyda, J.; Dzubiella, J.; Cremer, P. S.; Jungwirth, P. Beyond the Hofmeister series: Ion-specific effects on proteins and their biological functions. *J. Phys. Chem. B* **2017,** *121* (9), 1997-2014.
15. Nihonyanagi, S.; Yamaguchi, S.; Tahara, T. Counterion effect on interfacial water at charged interfaces and its relevance to the Hofmeister series. *J. Am. Chem. Soc.* **2014,** *136* (17), 6155-8.
16. Johnson, C. M.; Baldelli, S. Vibrational sum frequency spectroscopy studies of the influence of solutes and phospholipids at vapor/water interfaces relevant to biological and environmental systems. *Chem. Rev.* **2014,** *114* (17), 8416-46.
17. Jungwirth, P.; Cremer, P. S. Beyond Hofmeister. *Nat. Chem.* **2014,** *6* (4), 261-3.
18. Benjamin, I. Reaction dynamics at liquid interfaces. *Annu. Rev. Phys. Chem.* **2015,** *66*, 165-88.
19. Salis, A.; Ninham, B. W. Models and mechanisms of Hofmeister effects in electrolyte solutions, and colloid and protein systems revisited. *Chem. Soc. Rev.* **2014,** *43* (21), 7358-77.
20. Levin, Y.; dos Santos, A. P.; Diehl, A. Ions at the air-water interface: An end to a hundred-year-old mystery? *Phys. Rev. Lett.* **2009,** *103* (25), 257802.
21. Gurau, M. C.; Lim, S. M.; Castellana, E. T.; Albertorio, F.; Kataoka, S.; Cremer, P. S. On the mechanism of the hofmeister effect. *J. Am. Chem. Soc.* **2004,** *126* (34), 10522-3.
22. Darlington, A. M.; Jarisz, T. A.; DeWalt-Kerian, E. L.; Roy, S.; Kim, S.; Azam, M. S.; Hore, D. K.; Gibbs, J. M. Separating the pH-dependent behavior of water in the Stern and diffuse layers with varying salt concentration. *J. Phys. Chem. C* **2017,** *121* (37), 20229-20241.
23. Leontidis, E. Investigations of the Hofmeister series and other specific ion effects using lipid model systems. *Adv. Colloid. Interface. Sci.* **2017,** *243*, 8-22.

24. Doidge, E. D.; Carson, I.; Tasker, P. A.; Ellis, R. J.; Morrison, C. A.; Love, J. B. A Simple primary amide for the selective recovery of gold from secondary resources. *Angew. Chem. Int. Ed.* **2016,** *55* (40), 12436-9.
25. Tasker, P. A.; Plieger, P. G.; West, L. C. Metal complexes for hydrometallurgy and extraction, *Compr. Coord. Chem. II.* **2003**, 759-808.
26. Fowler, C. J.; Haverlock, T. J.; Moyer, B. A.; Shriver, J. A.; Gross, D. E.; Marquez, M.; Sessler, J. L.; Hossain, M. A.; Bowman-James, K. Enhanced anion exchange for selective sulfate extraction: Overcoming the Hofmeister bias. *J. Am. Chem. Soc.* **2008,** *130* (44), 14386-14387.
27. Uysal, A.; Rock, W.; Qiao, B.; Bu, W.; Lin, B. Two-step adsorption of $PtCl_6^{2-}$ complexes at a charged Langmuir monolayer: Role of hydration and ion correlations. *J. Phys. Chem. C* **2017,** *121* (45), 25377-25383.
28. Wen, Y. C.; Zha, S.; Liu, X.; Yang, S.; Guo, P.; Shi, G.; Fang, H.; Shen, Y. R.; Tian, C. Unveiling microscopic structures of charged water interfaces by surface-specific vibrational spectroscopy. *Phys. Rev. Lett.* **2016,** *116* (1), 016101.
29. Ma, G.; Chen, X.; Allen, H. C. Dangling OD confined in a Langmuir monolayer. *J. Am. Chem. Soc.* **2007,** *129* (45), 14053-7.
30. Du, Q.; Superfine, R.; Freysz, E.; Shen, Y. R. Vibrational spectroscopy of water at the vapor/water interface. *Phys. Rev. Lett.* **1993,** *70* (15), 2313-2316.
31. Hess, B.; Kutzner, C.; van der Spoel, D.; Lindahl, E. GROMACS 4: Algorithms for highly efficient, load-balanced, and scalable molecular simulation. *J. Chem. Theory. Comput.* **2008,** *4* (3), 435-47.
32. Best, R. B.; Zhu, X.; Shim, J.; Lopes, P. E.; Mittal, J.; Feig, M.; Mackerell, A. D., Jr. Optimization of the additive CHARMM all-atom protein force field targeting improved sampling of the backbone phi, psi and side-chain chi(1) and chi(2) dihedral angles. *J. Chem. Theory. Comput.* **2012,** *8* (9), 3257-3273.
33. Bjelkmar, P.; Larsson, P.; Cuendet, M. A.; Hess, B.; Lindahl, E. Implementation of the CHARMM Force field in GROMACS: Analysis of protein stability effects from correction maps, virtual interaction sites, and water models. *J. Chem. Theory. Comput.* **2010,** *6* (2), 459-66.
34. Lienke, A.; Klatt, G.; Robinson, D. J.; Koch, K. R.; Naidoo, K. J. Modeling platinum group metal complexes in aqueous solution. *Inorg. Chem.* **2001,** *40* (10), 2352-7.
35. Guardado-Calvo, P.; Atkovska, K.; Jeffers, S. A.; Grau, N.; Backovic, M.; Pérez-Vargas, J.; de Boer, S. M.; Tortorici, M. A.; Pehau-Arnaudet, G.; Lepault, J.; England, P.; Rottier, P. J.; Bosch, B. J.; Hub, J. S.; Rey, F. A. A glycerophospholipid-specific pocket in the RVFV class II fusion protein drives target membrane insertion. *Science* **2017,** *358* (6363), 663-667.
36. Cai, J.; Townsend, J. P.; Dodson, T. C.; Heiney, P. A.; Sweeney, A. M. Eye patches: Protein assembly of index-gradient squid lenses. *Science* **2017,** *357* (6351), 564-569.
37. Nagano, T.; Lubling, Y.; Várnai, C.; Dudley, C.; Leung, W.; Baran, Y.; Mendelson Cohen, N.; Wingett, S.; Fraser, P.; Tanay, A. Cell-cycle dynamics of chromosomal organization at single-cell resolution. *Nature* **2017,** *547*, 61.
38. Chen, S.; Itoh, Y.; Masuda, T.; Shimizu, S.; Zhao, J.; Ma, J.; Nakamura, S.; Okuro, K.; Noguchi, H.; Uosaki, K.; Aida, T. Subnanoscale hydrophobic modulation of salt bridges in aqueous media. *Science* **2015,** *348* (6234), 555-559.
39. González-Rubio, G.; Díaz-Núñez, P.; Rivera, A.; Prada, A.; Tardajos, G.; González-Izquierdo, J.; Bañares, L.; Llombart, P.; Macdowell, L. G.; Alcolea Palafox, M.; Liz-Marzán, L. M.; Peña-Rodríguez, O.; Guerrero-Martínez, A. Femtosecond laser reshaping yields gold nanorods with ultranarrow surface plasmon resonances. *Science* **2017,** *358* (6363), 640-644.

40. Sovago, M.; Vartiainen, E.; Bonn, M. Observation of buried water molecules in phospholipid membranes by surface sum-frequency generation spectroscopy. *J. Chem. Phys.* **2009,** *131* (16), 161107.
41. Chen, X.; Yang, T.; Kataoka, S.; Cremer, P. S. Specific ion effects on interfacial water structure near macromolecules. *J. Am. Chem. Soc.* **2007,** *129* (40), 12272-9.
42. Chen, X.; Flores, S. C.; Lim, S. M.; Zhang, Y.; Yang, T.; Kherb, J.; Cremer, P. S. Specific anion effects on water structure adjacent to protein monolayers. *Langmuir* **2010,** *26* (21), 16447-54.
43. Flores, S. C.; Kherb, J.; Cremer, P. S. Direct and reverse Hofmeister effects on interfacial water structure. *J. Phys. Chem. C* **2012,** *116* (27), 14408-14413.
44. Flores, S. C.; Kherb, J.; Konelick, N.; Chen, X.; Cremer, P. S. The effects of Hofmeister cations at negatively charged hydrophilic surfaces. *J. Phys. Chem. C* **2012,** *116* (9), 5730-5734.
45. Sung, W.; Wang, W.; Lee, J.; Vaknin, D.; Kim, D. Specificity and variation of length scale over which monovalent halide ions neutralize a charged interface. *J. Phys. Chem. C* **2015,** *119* (13), 7130-7137.
46. Raymond, E. A.; Tarbuck, T. L.; Brown, M. G.; Richmond, G. L. Hydrogen-bonding interactions at the vapor/water interface investigated by vibrational sum-frequency spectroscopy of HOD/H2O/D2O mixtures and molecular dynamics simulations. *J. Phys. Chem. B* **2003,** *107* (2), 546-556.
47. Nihonyanagi, S.; Ishiyama, T.; Lee, T. K.; Yamaguchi, S.; Bonn, M.; Morita, A.; Tahara, T. Unified molecular view of the air/water interface based on experimental and theoretical chi(2) spectra of an isotopically diluted water surface. *J. Am. Chem. Soc.* **2011,** *133* (42), 16875-80.
48. Bonn, M.; Nagata, Y.; Backus, E. H. G. Molecular structure and dynamics of water at the water-air interface studied with surface-specific vibrational spectroscopy. *Angew. Chem. Int. Ed.* **2015,** *54* (19), 5560-5576.
49. Scatena, L. F.; Brown, M. G.; Richmond, G. L. Water at hydrophobic surfaces: weak hydrogen bonding and strong orientation effects. *Science* **2001,** *292* (5518), 908-12.
50. Mondal, J. A.; Nihonyanagi, S.; Yamaguchi, S.; Tahara, T. Three distinct water structures at a zwitterionic lipid/water interface revealed by heterodyne-detected vibrational sum frequency generation. *J. Am. Chem. Soc.* **2012,** *134* (18), 7842-7850.
51. Tyrode, E.; Johnson, C. M.; Kumpulainen, A.; Rutland, M. W.; Claesson, P. M. Hydration state of nonionic surfactant monolayers at the liquid/vapor interface:  Structure determination by vibrational sum frequency spectroscopy. *J. Am. Chem. Soc.* **2005,** *127* (48), 16848-16859.
52. Livingstone, R. A.; Nagata, Y.; Bonn, M.; Backus, E. H. G. Two types of water at the water–surfactant interface revealed by time-resolved vibrational spectroscopy. *J. Am. Chem. Soc.* **2015,** *137* (47), 14912-14919.
53. Kropman, M. F.; Bakker, H. J. Dynamics of water molecules in aqueous solvation shells. *Science* **2001,** *291* (5511), 2118-20.
54. Ishihara, T.; Ishiyama, T.; Morita, A. Surface structure of methanol/water solutions via sum frequency orientational analysis and molecular dynamics simulation. *J. Phys. Chem. C* **2015,** *119* (18), 9879-9889.
55. Ishiyama, T.; Terada, D.; Morita, A. Hydrogen-bonding structure at zwitterionic lipid/water interface. *J. Phys. Chem. Lett.* **2016,** *7* (2), 216-20.

56. Nagata, Y.; Mukamel, S. Vibrational sum-frequency generation spectroscopy at the water/lipid interface: Molecular dynamics simulation study. *J. Am. Chem. Soc.* **2010,** *132* (18), 6434-6442.
57. Sovago, M.; Campen, R. K.; Wurpel, G. W. H.; Müller, M.; Bakker, H. J.; Bonn, M. Vibrational response of hydrogen-bonded interfacial water is dominated by intramolecular coupling. *Phys. Rev. Lett.* **2008,** *100* (17), 173901.
58. Gan, W.; Wu, D.; Zhang, Z.; Feng, R. R.; Wang, H. F. Polarization and experimental configuration analyses of sum frequency generation vibrational spectra, structure, and orientational motion of the air/water interface. *J. Chem. Phys.* **2006,** *124* (11), 114705.
59. Wang, H.-F.; Gan, W.; Lu, R.; Rao, Y.; Wu, B.-H. Quantitative spectral and orientational analysis in surface sum frequency generation vibrational spectroscopy (SFG-VS). *Int. Rev. Phys. Chem.* **2005,** *24* (2), 191-256.
60. Knight, A. W.; Qiao, B.; Chiarizia, R.; Ferru, G.; Forbes, T.; Ellis, R. J.; Soderholm, L. Subtle effects of aliphatic alcohol structure on water extraction and solute aggregation in biphasic water/n-dodecane. *Langmuir* **2017,** *33* (15), 3776-3786.
61. Rock, W.; Oruc, M. E.; Ellis, R. J.; Uysal, A. Molecular scale description of anion competition on amine-functionalized surfaces. *Langmuir* **2016,** *32* (44), 11532-11539.

**TOC GRAPHICS**

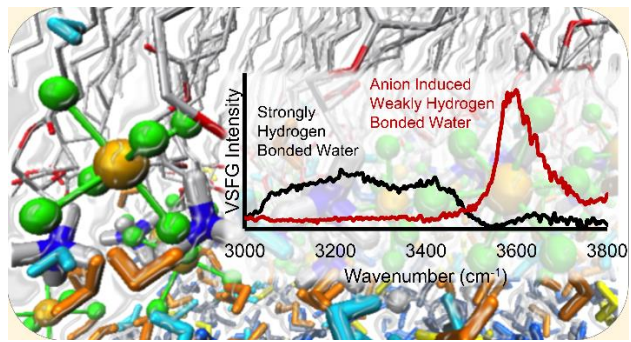